%% file: main.tex
\documentclass[sigconf]{acmart}
\AtBeginDocument{%
  }

\setcopyright{acmlicensed}
\copyrightyear{2025}
\acmYear{2025}
\acmDOI{XXXXXXX.XXXXXXX}
\acmConference[Conference acronym 'XX]{Make sure to enter the correct
  conference title from your rights confirmation email}{June 03--05,
  2018}{Woodstock, NY}
\acmISBN{978-1-4503-XXXX-X/2018/06}

\usepackage{graphicx}
\usepackage{amsthm}
\usepackage{amsmath}
\usepackage{dsfont}
\usepackage{graphicx} 
\usepackage{booktabs} 

\usepackage{enumitem}
\usepackage{amsfonts}
\usepackage{cleveref} 
\usepackage{algorithm}
\usepackage{algpseudocode}
\usepackage{amsmath}
\usepackage[most]{tcolorbox}
\usepackage{enumitem}

\begin{document}

\title{LESER: Learning to Expand via Search Engine-feedback Reinforcement in e-Commerce}


\author{Yipeng Zhang}
\email{yipezhang@ebay.com}
\affiliation{%
  \institution{eBay Inc.}
    \city{San Jose}
  \state{CA}
  \country{USA}
}
\author{Bowen Liu}
\email{boweliu@ebay.com}
\affiliation{%
  \institution{eBay Inc.}
  \city{Shanghai}
  \country{China}
}

\author{Xiaoshuang Zhang}
\email{xiaoszhang@ebay.com}
\affiliation{%
  \institution{eBay Inc.}
  \city{Shanghai}
  \country{China}
}

\author{Aritra Mandal}
\email{arimandal@ebay.com}
\affiliation{%
  \institution{eBay Inc.}
    \city{San Jose}
  \state{CA}
  \country{USA}
}

\author{Canran Xu}
\email{canxu@ebay.com}
\affiliation{%
  \institution{eBay Inc.}
  \city{Shanghai}
  \country{China}
}

\author{Zhe Wu}
\email{zwu1@ebay.com}
\affiliation{%
  \institution{eBay Inc.}
  \city{San Jose}
  \state{CA}
  \country{USA}
}

\renewcommand{\shortauthors}{Zhang et al.}

\begin{abstract}

\input{section/0_abstract}

\end{abstract}

\begin{CCSXML}
<ccs2012>
<concept>
       <concept_id>10002951.10003317.10003325.10003330</concept_id>
       <concept_desc>Information systems~Query reformulation</concept_desc>
       <concept_significance>500</concept_significance>
       </concept>
   <concept>
       <concept_id>10010147.10010257.10010258.10010261</concept_id>
       <concept_desc>Computing methodologies~Reinforcement learning</concept_desc>
       <concept_significance>500</concept_significance>
       </concept>
   
 </ccs2012>
\end{CCSXML}

\ccsdesc[500]{Computing methodologies~Reinforcement learning}
\ccsdesc[500]{Information systems~Query reformulation}

\keywords{Query expansion, Reinforcement learning;  Large language models, E-commerce search}


\maketitle

\input{section/1_introduction}

\input{section/2_relatedworks}

\input{section/3_motivation}

\input{section/3_5_method}

\input{section/4_results}
\input{section/5_conclusion}

\bibliographystyle{ACM-Reference-Format}
%\bibliography{sample-base}
\bibliography{acmart}

\appendix
\input{section/6_appendix}

\end{document}

%% file: section/0_abstract.tex
User queries in e-commerce search are often vague, short, and underspecified, making it difficult for retrieval systems to match them accurately against structured product catalogs. This challenge is amplified by the one-to-many nature of user intent, where a single query can imply diverse and competing needs. Existing methods, including neural query expansion and prompting-based LLM approaches, fall short in real-world settings: they struggle to capture nuanced user intent, often generate outputs that violate platform constraints, and rely on workflows that are difficult to scale in production. We propose \underline{L}earning to \underline{E}xpand via \underline{S}earch \underline{E}ngine-feedback \underline{R}einforcement (\texttt{LESER}), a novel framework that fine-tunes a context-aware LLM using real-time search engine feedback as supervision. LESER formulates query expansion as a retrieval optimization task and leverages Group Relative Policy Optimization to learn directly from relevance and coverage metrics. LESER is trained to reason over search results and produce high quality query expansions that align with platform rules and retrieval objectives. We evaluate LESER on large-scale, real-world e-commerce datasets, demonstrating substantial improvements in both offline and online settings. Our results show that LESER not only enhances semantic coverage and retrieval relevance but also delivers measurable gains in user engagement, making it a practical and scalable solution for modern search systems.

%% file: section/1_introduction.tex
\section{Introduction}

In the modern age of conversational AI assistants and voice-driven searches, such as ChatGPT~\cite{achiam2023gpt} and Alexa~\cite{kumar2017just}, users increasingly rely on succinct and imprecise queries. Queries such as \textit{"sleeping pill"} or \textit{"baby monitor with night vision"} often omit key product-defining details, presenting a significant retrieval challenge. This mismatch between vague user language and product specifications frequently yields irrelevant or incomplete results, diminishing user engagement and causing potential revenue loss. This difficulty stems from the fundamental \textbf{one-to-many mapping} challenge in modern query expansion. For example, the query \textit{"warm winter coat"} could mean a user wants a \textit{highly insulated down parka}, a \textit{waterproof rain shell}, or a \textit{stylish wool overcoat}. A retrieval system must understand these different needs to show relevant and comprehensive products.

\input{./fig/figure1}

Traditional query expansion techniques struggle with this complexity, as they typically rely on augmenting queries with semantically related terms~\cite{carpineto2012survey}. A keyword-based method, for example, might add "support" to a query for \textit{"back pain relief"}, but it cannot infer and expand to product specifications that a user might actually want. Later methods used more advanced neural models to improve semantic matching and recall~\cite{roy2016using, naseri2021ceqe}. However, these neural models focus only on surface-level semantic understanding, lacking the domain-specific knowledge to translate a user's broad intent into a set of effective, rewritten queries, thus failing to improve retrieval relevance.

Recent approaches \cite{jagerman2023query,wang2023query2doc} have explored using large language models (LLMs) to perform context-aware query expansion by incorporating retrieved results. However, these methods often rely on in-context learning or multi-step agentic workflows \cite{zhang2024query}, which can be inefficient and difficult to scale in production environments. Moreover, despite their generative capabilities, LLMs~\cite{Touvron2023LLaMA,Yang2025Qwen3} frequently produce hallucinated outputs that violate platform-specific constraints or business rules~\cite{huang2025survey}, undermining their reliability in e-commerce applications without proper finetuning.

To address the aforementioned challenges, we propose \underline{L}earning to \underline{E}xpand via \underline{S}earch \underline{E}ngine-feedback \underline{R}einforcement (\texttt{LESER}), a novel framework that performs query expansion using a context-aware LLM trained through search-engine-in-the-loop reinforcement learning (RL). As illustrated in \Cref{fig:fig1}, \texttt{LESER} leverages the retrieval set from the raw user query as contextual grounding, enabling domain-aware reasoning to generate expansions informed by real-time search engine information. To handle the inherent one-to-many mapping challenge in query expansion, \texttt{LESER} uses real-time search engine feedback, specifically relevance and retrieval size, as reward signals, thereby eliminating the need for supervised labeling during training. To effectively optimize the LLM toward these rewards, we adopt Group Relative Policy Optimization (GRPO)~\cite{shao2024deepseekmath}, which selectively rewards reasoning steps and expansions that yield superior real-world performance. This search-engine-in-the-loop RL paradigm enables \texttt{LESER} to learn directly from live retrieval signals, ensuring that each generated expansion is grounded in actual catalog content and optimized for both comprehensive retrieval and relevance.

We evaluate LESER on a large real-world e-commerce dataset spanning a diverse range of query types. Our experiments demonstrate that LESER significantly improves semantic coverage and retrieval effectiveness compared to strong baseline approaches. Online A/B testing confirms its scalability and production-readiness, yielding measurable improvements in relevance, diversity, and click-through rates. Our contributions are summarized as follows:

\begin{itemize}[leftmargin=*]
    \item We propose \texttt{LESER}, a search-engine-in-the-loop reinforcement learning framework for optimizing e-commerce query expansion. \texttt{LESER} leverages a LLM to perform domain-aware reasoning based on contextual information retrieved from the search engine.
    \item \texttt{LESER} introduces a novel reward formulation that leverages real-time search engine feedback as direct optimization signals to effectively guides the LLM to generate high-quality query expansions without requiring annotations, ensuring alignment with real-world search performance.
    \item We demonstrate the effectiveness of \texttt{LESER} through comprehensive offline and online evaluations, highlighting its real-world impact on retrieval quality and user experience.
\end{itemize}

%% file: fig/figure1.tex
\begin{figure*}[t]
    \centering
    \includegraphics[width=\textwidth]{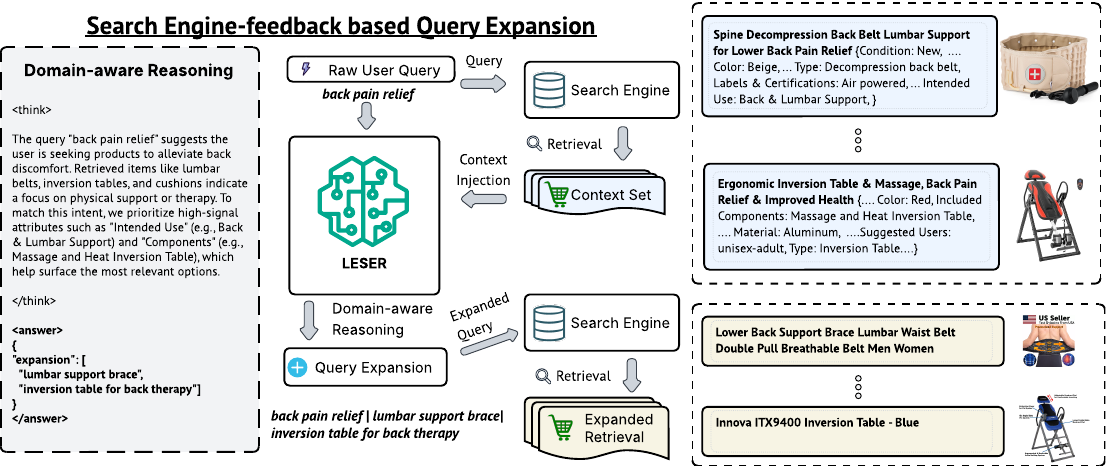}
    \caption{\textbf{LESER Overview.}
Given a user query, a context set is retrieved from the search engine. The LLM then performs domain-aware reasoning, trained via search-engine-in-the-loop RL, to generate query expansions, resulting in relevant and more diverse search results (The reasoning part is simplified due to the space constraints).}
    \label{fig:fig1}
\end{figure*}

%% file: section/2_relatedworks.tex
\section{Related Works}

\paragraph{Query Expansion.}
Classic query expansion focused on term-level augmentation like Rocchio feedback or synonym injection to address lexical gaps~\citep{Rocchio1971Relevance,Lavrenko2001RM3,mandal2019query}. Subsequent neural approaches introduced more sophisticated signals, using paraphrasing to improve semantic matching~\citep{Nogueira2019Doc2query,Nogueira2020Expando}, dense-retrieval PRF~\citep{Wang2021PRFMRDR}, or user interactions~\citep{Mandal2023Semantic}. More recently, reinforcement learning has been used to optimize rewrites against click metrics~\citep{nguyen2025rl}. Our work builds on this trajectory but differs critically by using a rich set of search feedback signals to optimize for diverse expansions.

\paragraph{LLMs for e-Commerce.}
General-purpose LLMs underperform in commerce search as they lack intrinsic knowledge of a platform's product catalog and business rules~\citep{PalenMichel2024Investigating}. Adaptation techniques are therefore necessary, ranging from Retrieval-Augmented Generation (RAG) in models like DRaGAN~\citep{Zhao2023Dragan} to using LLMs for explainable relevance learning~\citep{zhao2025explainable}. However, these approaches predominantly depend on the inherent capabilities of pretrained models, which are not explicitly optimized for domain-specific adaptation or alignment with e-commerce specific logic.

\paragraph{Generative Retrieval.}
A distinct paradigm, generative retrieval, reframes retrieval as a sequence generation task where models like the Differentiable Search Index (DSI) directly generate document identifiers~\citep{Tay2022DSI,Mehta2022DSIpp}. Subsequent work has focused on addressing practical challenges such as incremental updates, sharding, and e-commerce alignment~\cite{Pang2025GRAM,Huynh2024PromptDSI,Neague2024DeDSI}. However, significant open problems in scaling to million-SKU catalogs and ensuring synchronization with dynamic inventory persist.

%% file: section/3_motivation.tex
\section{Motivation}
\subsection{Heterogeneous User Queries}

User queries in e-commerce settings are often ambiguous, underspecified, or colloquial. Many users initiate searches with vague or partial expressions due to limited product knowledge or uncertainty about what they need. For example, a query like \textit{“sleeping pill”} conveys general intent but lacks specificity about the active ingredient, which could be melatonin, diphenhydramine, or doxylamine. Similarly, \textit{“phone charger”} omits essential distinctions such as connector type (USB-C, Lightning, MagSafe) or required wattage. A query like \textit{“red dress for wedding”} implies intent (event-appropriate attire) but leaves out important filters like fabric, cut, or length—attributes that are inconsistently labeled across consumer-to-consumer (C2C) listings. Even more indirectly, a user searching for \textit{“kids shoes with wheels”} may actually intend to find Heelys, a brand name that may not appear in the title or metadata. In these cases, resolving user intent requires leveraging internal product taxonomies, language model priors, and structured knowledge graphs to bridge vague surface forms with precise product representations.

This problem is exacerbated in C2C marketplaces, where inventory is uncurated and seller-provided metadata is often sparse or noisy. For instance, identical clothing items may be labeled as \textit{“2XL”} and \textit{“XXL”}, or a used phone might state \textit{“no scratches”} in free-form text instead of using a standardized condition field like \textit{“Used - Very Good”}. In such scenarios, surface-level matching is insufficient, and systems must perform deeper semantic alignment between user queries and listing-specific attributes. The increasing adoption of large language models (LLMs) has further shifted user behavior toward more conversational and exploratory queries, raising the bar for retrieval systems to accurately infer latent user intent and align it with structured item data.

\subsection{Limits of Conventional Search}

Traditional search pipelines typically consist of modular components, including query understanding (e.g., named entity recognition and intent classification), embedding-based retrieval, and learning-to-rank models. While effective for well-specified queries, these systems face critical limitations when handling underspecified, ambiguous, or structurally complex inputs. A particularly challenging case involves alphanumeric identifiers such as ISBNs for books, OEM part numbers for auto parts, or model codes for electronics (e.g., \textit{“WD19TB”} for a Dell docking station). These tokens are essential for product disambiguation but may be inconsistently represented across item titles or metadata fields. Embedding-based retrieval often struggles with such tokens, as their latent spaces are not designed to preserve structured symbol composition. Meanwhile, token-level exact match is brittle, failing under formatting inconsistencies (e.g., \textit{“WD-19TB”} vs. \textit{“WD19 TB”}) or missing substrings.

The problem intensifies when users provide only partial information and expect the system to fill in the gaps. For example, a query like \textit{“replacement battery for Canon camera”} lacks a model number but presumes the system can infer compatible SKUs from historical co-occurrence or product knowledge. Conventional pipelines, which operate with limited cross-module coordination and shallow intent modeling, struggle to resolve such cases. These limitations underscore the need for intent-centric retrieval systems capable of performing reasoning over heterogeneous signals—query text, retrieved context, structured attributes, and domain knowledge.

\subsection{The Role of Modern Language Models}

Recent advances in large language models (LLMs), particularly in retrieval-augmented generation (RAG) and tool-augmented agents, offer promising solutions to the limitations above. These models can dynamically enrich under-specified queries, infer latent constraints, and generate structured, schema-aligned representations grounded in product taxonomies. By leveraging retrieved context and internal knowledge, LLMs can bridge the gap between user-facing fuzzy intent and machine-consumable structured queries—forming the foundation for more interpretable, effective, and efficient e-commerce search.

%% file: section/3_5_method.tex
\section{Method}

\subsection{Context-Grounded Query Expansion}
To ground query expansions in real catalog context, the LESER framework is inspired from classic Pseudo-Relevance Feedback (PRF) technique~\cite{Rocchio1971Relevance, li2025llm}, treating the top-$k$ items from an initial search as a pseudo-relevant context set. Unlike traditional PRF methods that rely on simple keyword statistics, LESER leverages the full spectrum of product data by processing both the unstructured text of item titles and its rich, structured attribute-value pairs. For instance, as shown in \Cref{fig:fig1}, it extracts attributes that define a product’s core function, like \{\textit{"Intended Use": "Back \& Lumbar Support"}\}, or its specific category, like \{\textit{"Type": "Inversion Table"}\}, rather than less informative details. 
%In practice, query expansions in search systems typically fall into two categories: (1) free-form expansions, which enrich the query with additional descriptive or intent-expressive language (e.g., “lightweight waterproof jacket for hiking”), and (2) schema-aligned expansions, which map inferred intent to platform-compatible concepts such as categories, brands, and features. 
LESER is designed to support both forms simultaneously, allowing the LLM to produce diverse, complementary refinements that enhance both retrieval quality and diversity. The prompt used to guide the LLM to generate query expansions is shown in~\Cref{tab:grpo_full_prompt}. More specifically, the query expansion process performed by the LLM includes a context-aware reasoning step, encapsulated within the \texttt{<think>}...\texttt{</think>} tokens. After completing the reasoning, the model is expected to generate the final query expansion inside the \texttt{<answer>}...\texttt{</answer>} tokens, formatted as a JSON string. The expanded retrieval set is defined as the union of the results retrieved by the original query and all corresponding expansions under the "expansion" field in the JSON object.

\input{./tab/tab1}
%\input{./algo/algo}
% \subsection{Policy Optimization with Search Engine Feedback}
\subsection{Optimization with Search Feedback}
\paragraph{Search Engine Feedback as Reward.}
\label{sec:reward_design}

A central challenge of context-grounded query expansion is the absence of ground-truth labels to supervise the training process. To address this, \texttt{LESER} optimizes for query expansion leveraging real-time feedback from the search engine to guide and reward high-quality expansions. Specifically, we frame the evaluation of query expansion as a retrieval task, using the search engine itself as an oracle for feedback. The core intuition is that a high-quality expansion should retrieve a set of items that is both relevant to the original query and diverse in its retrieval coverage. This principle allows us to define a reward function based on search feedback and optimize the LLM directly against it.

Formally, let \( q \) denote one original user query. We define one expansion set as:
$\mathcal{Y} = \{q, \hat{y}_1, \hat{y}_2, \ldots, \hat{y}_M\}$
where each \( \hat{y}_m \) is one expansion with in the "expansion" field. The reward is defined over the entire set \( \mathcal{Y} \), including the original query \( q \), and reflects the quality of the combined retrieval result on the relevance metric $r_\text{rel}(\mathcal{Y})$ and recall size metric $r_\text{size}(\mathcal{Y})$.

The overall reward function is defined as:
\begin{equation}\label{eq:reward_func}
r(\mathcal{Y}) = 
\begin{cases}
r_\text{rel}(\mathcal{Y}) + \lambda \cdot r_\text{size}(\mathcal{Y}) & \text{if } \mathcal{Y} \text{ is valid} \\
0 & \text{otherwise}
\end{cases}
\end{equation}
where the two components are defined as:
\begin{equation} \label{eq:reward_components}
\begin{split}
  r_\text{rel}(\mathcal{Y}) = \frac{Rel(\mathcal{Y})}{Rel(q) + \epsilon}, \text{ }
  r_\text{size}(\mathcal{Y}) = \frac{Ret(\mathcal{Y})}{Ret(q) + \epsilon}.
\end{split}
\end{equation}
Here, \( Rel(\mathcal{Y}) \) denotes the average relevance score of top-$k$ items retrieved using all queries in \( \mathcal{Y} \), and \( Ret(\mathcal{Y}) \) is the total number of unique items retrieved across the expansion set. The terms \( Rel(q) \) and \( Ret(q) \) represent the same metrics computed using the original query \( q \) alone, serving as a baseline for normalization. The hyperparameter \( \lambda \) controls the trade-off between relevance and retrieval breadth, and \( \epsilon \) is a small constant (\( 10^{-4} \)) to prevent division by zero. Furthermore, since the $\mathcal{Y}$ is generated by an LLM trained to perform context-aware reasoning, the output must follow a predefined format: a reasoning section enclosed within \texttt{<think>}...\texttt{</think>} tokens, followed by a set of query expansions enclosed within \texttt{<answer>}...\texttt{</answer>} tokens and formatted as a JSON object. To ensure compatibility with downstream retrieval systems, we enforce a strict validation gate by assigning a reward of zero to any structurally invalid output. An output is deemed invalid for reasons including malformed syntax, unsupported attributes, or excessive length. This policy acts as a crucial filter, guaranteeing that the learning process is guided exclusively by well-formed and executable expansions.

\paragraph{Group Relative Policy Optimization (GRPO)}
Given the one-to-many nature of query expansion, where multiple expansion \emph{sets} may be equally valid, we adopt GRPO~\cite{shao2024deepseekmath} to fine-tune the model using relative rewards across candidates. For each input \(x\) (original query \(q\) with its item context), the policy \(\pi_\theta\) samples \(N\) candidate \emph{sets} of expansions at a high sampling temperature to encourage exploration:
$\big\{\hat{\mathcal{Y}}^{(i)}\big\}_{i=1}^{N},
\hat{\mathcal{Y}}^{(i)}=\{\hat{y}^{(i)}_{1},\ldots,\hat{y}^{(i)}_{K}\}$,
where \(K\) is the number of expansions in each set. The full expansion set used for retrieval is
\(\mathcal{Y}^{(i)}=\{q\}\cup \hat{\mathcal{Y}}^{(i)}\). Each \(\mathcal{Y}^{(i)}\) is submitted to the search engine to obtain a result set, from which we compute a scalar reward
\(
r^{(i)} = r(\mathcal{Y}^{(i)}),
\)
as defined in~\Cref{eq:reward_func}. The group-relative advantage for candidate \(i\) is then $A^{(i)} = r^{(i)} - \frac{1}{N}\sum_{j=1}^{N} r^{(j)}$. Since each expansion set \(\hat{\mathcal{Y}}^{(i)}\) is serialized into a single output sequence (with expansions separated by special tokens), the likelihood of generating the set is simply the sequence likelihood under the model. We therefore define the probability ratio $\rho^{(i)} = \frac{\pi_\theta(\hat{\mathcal{Y}}^{(i)} \mid x)}{\pi_{\theta_{\text{old}}}(\hat{\mathcal{Y}}^{(i)} \mid x)}$,
where both numerator and denominator are computed as products of token-level probabilities over the entire serialized sequence. Finally, the GRPO objective combines clipped policy optimization with a KL regularizer:
\begin{align}
\mathcal{L}_{\text{GRPO}}
&= -\,\mathbb{E}_{x}
\left[
\frac{1}{N}\sum_{i=1}^N
\min \!\left(
\rho^{(i)} A^{(i)},\;
\operatorname{clip}\!\big(\rho^{(i)},\,1-\epsilon,\,1+\epsilon\big) A^{(i)}
\right)
\right] \nonumber \\
&\quad + \; \beta \,\mathbb{E}_{x}
\left[
\frac{1}{N}\sum_{i=1}^N
\Big( r_{\text{KL}}^{(i)} - \log r_{\text{KL}}^{(i)} - 1 \Big)
\right].
\label{eq:grpo-loss}
\end{align}

In this objective, \(\epsilon\) specifies the clipping range for stabilizing updates, while \(\beta\) controls the strength of the KL regularization. The KL ratio for candidate \(i\) is defined as
\(
r_{\text{KL}}^{(i)} \;=\;
\frac{\pi_\theta\!\big(\hat{\mathcal{Y}}^{(i)} \mid x\big)}
     {\pi_{\theta_{\text{old}}}\!\big(\hat{\mathcal{Y}}^{(i)} \mid x\big)},
\)
which compares the likelihood of the same expansion set under the current and reference policies.

\paragraph{Supervised Warm-Up and GRPO Training.}
Before GRPO training, we perform a supervised fine-tuning warm-up phase to stabilize reinforcement learning and prevent the model from generating random or syntactically invalid outputs that hinder learning efficiency. Specifically, we first train an SFT model on 10,000 examples distilled from a general in-house dataset, where GPT-4.1 is prompted to produce both reasoning and a final answer in the format \texttt{<think> Reasoning </think>} and \texttt{<answer> Answer </answer>}. The reasoning section explicitly explains why the answer is appropriate, injecting domain-specific business logic and encouraging adherence to the structured response format. This supervised warm-up is run for one epoch. We then continue with GRPO training, which differs from SFT in requiring interaction with a search engine to compute rewards. For this stage, we use a dataset of 100,000 real-world search queries, where each generated structured intent is executed against the search engine, and the resulting retrieval set is used to calculate the reward.

%% file: tab/tab1.tex
\begin{table}[t]
\centering
\small
\renewcommand{\arraystretch}{1.05}  % Slightly tighter line spacing
\setlength{\tabcolsep}{5pt}         % Reduce horizontal padding between columns
\begin{tabular}{p{8cm}}
\toprule
\textbf{System Prompt:} \texttt{You are a helpful and safe assistant. In your responses, include a <think> </think> section for reasoning and an <answer> </answer> section for the final answer.} \\
\midrule
\textbf{Instruction} \texttt{
Given a user query <query> \{user query\}</query> and a list of retrieved items, where each item includes a title and a set of attribute name–value pairs \{item details\}, your task is to:
1. Infer the user's intent behind the query.  
2. Identify and select the most relevant attribute-value pairs that can help clarify, disambiguate, or enrich the query.  
3. If the user's intent is already specific and unambiguous, do not generate any expansions—return an empty list.  
4. Otherwise, generate multiple diverse query expansions (at least three), each incorporating different combinations of relevant attribute-value pairs to improve item retrieval.  
5. Ensure that each expansion remains faithful to the original intent while offering complementary refinements.  
6. Return a JSON object in the following format, replacing "expanded\_query" with your generated queries:  
\{"expansion": ["expanded\_query 1", "expanded\_query 2", "expanded\_query 3"]\}  
If no expansion is needed, return: \{"expansion": []\}.
} \\ % <--- THIS IS THE FIX
\bottomrule
\end{tabular}
\caption{Prompt used for domain context-aware query expansions.}
\label{tab:grpo_full_prompt}
\end{table}

%% file: section/4_results.tex
\section{Results}
\subsection{Implementation Details}
\paragraph{Hyperparameter.} We adopt a two-stage training pipeline to align large language models with structured query generation tasks in e-commerce. All experiments are conducted using the Ms-Swift framework ~\cite{zhao2024swiftascalablelightweightinfrastructure}. In the supervised fine-tuning (SFT) warm up stage, we perform full-parameter updates using \texttt{bfloat16} precision, a global batch size of 64 (with per-device batch size of 16 and gradient accumulation), a learning rate of $1 \times 10^{-7}$, and a warmup ratio of 5\%. DeepSpeed ZeRO-3 \cite{rajbhandari2020zero} is employed for memory-efficient optimization. In the GRPO training, the batch size is set to $8$ with no gradient accumulation. During the training, the model generates $N = 4$ candidates per input using temperature sampling ($\tau = 0.9$), and is optimized with a learning rate of $1 \times 10^{-6}$ with reward weight $\lambda = 0.1$ and $\beta = 0.04$. Both stages are trained on 8 NVIDIA H200 GPUs using full-parameter updates, with a maximum input length of $2048$. Furthermore, in constructing the instruction, the context set is chosen to be no more than 10 items. In the reward calculation, the relevance score is generated from internal relevance model from retrieved top 10 items.

\paragraph{Relevance Model.} Following the approach in~\cite{saha2025improving}, the relevance model is based on a Gradient Boosted Decision Tree (GBDT) trained in a pointwise manner using Gaussian regression. It uses human-judged relevance labels on a five-point scale (Perfect, Excellent, Good, Fair, Bad) and incorporates incremental updates to handle new sponsored listings. The model outputs a continuous relevance score for each item comparing to the query.
\subsection{Quantitative Evaluation}
\paragraph{Evaluation Metric.}  
We directly evaluate the effectiveness of the proposed approach on a dataset of 10,000 real-world e-commerce queries. The quality of query expansions is assessed by interacting with a production search engine. For a given original query, let $\mathcal{R}_{\text{orig}}$ be its set of retrieved items and $\mathrm{AvgRel}_{\text{orig}}$ be the average relevance score of its top 100 results. Similarly, let $\mathcal{R}_{\text{exp}}$ denote the union of items retrieved from both the original query and its corresponding expansions, with $\text{AvgRel}_{\text{exp}}$ as its associated average relevance score.
Then, we define two primary metrics to measure the impact of the expansions. (1) \textbf{Retrieval Gain}, defined as $\Delta\mathsf{Ret} = |\mathcal{R}_{\text{exp}} \setminus \mathcal{R}_{\text{orig}}|$, which quantifies the increase in the number of unique retrieved items; and (2) \textbf{Relevance Gain}, defined as 
%$\Delta\mathsf{Rel} = \frac{\mathrm{AvgRel}_{\text{exp}} - \mathrm{AvgRel}_{\text{orig}}}{\mathrm{AvgRel}_{\text{orig}}}$
$\Delta\mathsf{Rel} = \mathrm{AvgRel}_{\text{exp}} - \mathrm{AvgRel}_{\text{orig}}$, 
and measures the improvement in relevance introduced by query expansion. For our analysis, we report the percentage of queries that exhibit a positive gain for each metric. To ensure a robust evaluation, we repeat the generation process five times for each query using a temperature of 0.5, reporting the mean and standard deviation across these runs. 

\paragraph{Baseline Models.}
To assess the effectiveness of \texttt{LESER} in query expansion, we compare our proposed approach against three strong baselines. 
(1) \textit{Embedding-based Query Expansion (EBQE)} approach~\cite{naseri2021ceqe, Mandal2023Semantic}, a non-LLM approach that encodes the input query to retrieve semantically similar queries from search logs, and then select the expansion based on the embedding similarities.
(2) \textit{GPT-4.1 with Few-Shot Prompting}: Following the recent methodologies for LLM-driven query expansion, we employ a few-shot prompting strategy similar as proposed in ~\cite{lei2024corpus}. Specifically, the model is given our \texttt{LESER} instruction prompt along with three in-context examples to perform the task. (3) \textit{Supervised Fine-Tuning (SFT-LLM)} model, which uses the same LLM backbone as \texttt{LESER}. Inspired by the industrial methodology in~\citet{peng2024beque}, we construct the training data using a multi-stage rejection sampling pipeline. This pipeline filters candidate rewrites for semantic relevance and retrieval increment to improve search results.
% This strong supervised baseline allows us to precisely isolate and measure the additional value provided by our reinforcement learning (GRPO) over standard supervised adaptation.

% To fairly assess the effectiveness of LESER in query expansion, we compare our proposed approach against three strong baselines: 
% (1) \textit{Embedding-based Query Expansion}, similar to approaches in~\cite{mandal2019query, naseri2021ceqe}, which performs query expansion by identifying historically similar queries using embeddings. It encodes the input query and retrieves past queries with the highest embedding similarity from user interaction logs. The final expansion terms are then selected based on a historical performance threshold.
% (2) \textit{GPT-4.1 with Few-Shot Prompting (GPT4.1)}, which uses the same instruction format as the GRPO training prompt (see~\Cref{tab:grpo_full_prompt}), along with domain-specific guidance and three in-context examples. This allows for a direct comparison between instruction-tuned learning and inference-time prompting; and 
% (3) \textit{Supervised Fine-Tuning (SFT-LLM)}, where the same LLM architecture as LESER is trained on labeled query expansions collected from historical user data, using the identical prompt as in our method. This baseline assesses the value of reinforcement learning (GRPO) in a problem setting with multiple valid answers.

\paragraph{Performance Comparison.}
% \yipeng{I want to change here}

As shown in \Cref{tab:exp1}, \texttt{LESER}, built on a LLaMA 3.2-3B-Instruct, outperforms all baseline models across both retrieval and relevance metrics, with results averaged over five independent runs. It demonstrates a strong ability to expand retrieval coverage, retrieving additional relevant items for over 70\% of queries, while also improving semantic relevance. In contrast, traditional embedding-based method (EBQE) exhibits zero variance due to the deterministic nature and fail to significantly improve retrieval or relevance, likely because they lack awareness of the retrieval context during expansion. While GPT-4.1 serves as a strong few-shot baseline, it relies solely on prompting and lacks long-term adaptation. \texttt{LESER} achieves consistently higher gains by learning directly from interaction with the search engine. The SFT-LLM baseline shows low performance variance, but this stability comes at a cost: learning from static, deterministic labels limits its generalization ability and leads to rigid, overfitted generation patterns that harm retrieval quality. In contrast, the slightly higher standard deviation in \texttt{LESER} reflects the model's exploration of diverse, high-reward expansion strategies encouraged by the GRPO framework. By incorporating real-time feedback during training, \texttt{LESER} internalizes domain-specific patterns and learns to generate expansions that are both broad in coverage and grounded in relevance. %This dynamic, feedback-driven learning paradigm is central to its robust and superior performance in real-world search scenarios.

\input{./tab/tab2}

\paragraph{Ablation Studies.}
\input{./tab/tab3}
 We conduct ablation studies in \Cref{tab:exp2} to evaluate the effectiveness of our design choices and the generalization capability of our proposed framework. These studies examine the impact of using different backbone models including LLaMA 3.1-8B-Instruct~\cite{meta_llama31} and LLaMA 3.2-1B-Instruct~\cite{meta_llama32}, as well as variations in the training paradigm, including SFT-based warm-up and prompting a pretrained LLaMA 3.2-3B-Instruct. Although the larger LLaMA 3.1-8B-Instruct backbone achieves performance comparable to the 3B model, we choose for the latter due to latency constraints in potential real-time online deployments. In contrast, the smaller LLaMA 3.2-1B-Instruct backbone struggles to follow prompts reliably and fails to perform the task effectively, highlighting its limited capacity. We also observe that models trained with GRPO alone (without warm-up) could fail to follow the system prompt during training. In particular, they may omit the required \texttt{<think>} and \texttt{<answer>} tokens, preventing the reward function, relying on the \texttt{<answer>} block, from correctly extracting and evaluating the final prediction. This leads to inefficient training and degraded performance. Interestingly, we observe similar behavior when experimenting with alternative instructions during fine-tuning of larger backbones (3B and 8B). Inappropriately designed prompts can result in the model ignoring system instructions, which breaks alignment with the reward function. Our observation suggests that carefully adjusting the prompt format was essential to ensure that the model generated structured and meaningful outputs, enabling the reward function to operate effectively.

\subsection{Qualitative Evaluation}
\label{sec:qualitative_eval}
As shown in \Cref{tab:qualitative_examples}, \texttt{LESER} demonstrates a powerful capability to transform ambiguous searches by leveraging its understanding of the product catalog. This adaptability is evident across different query types. For instance, \texttt{LESER} refrains from expanding a highly specific query like "\textit{black bose quietcomfort ultra headphones}," while for a broader term such as "\textit{ergonomic mouse}," it generates distinct queries like "\textit{wireless vertical mouse for wrist support}" to guide users toward different product forms. The true strength of \texttt{LESER} is revealed with vague, problem-oriented needs like "\textit{sleep improvement}," which typically fail on standard e-commerce platforms. In such cases, \texttt{LESER} deconstructs the general problem into multiple actionable searches across entirely different solution categories, suggesting pathways from supplements ("\textit{natural sleep aid supplement}") and external aids ("\textit{herbal sleep relief patch}") to physical devices ("\textit{anti-snoring device}") and electronic monitors ("\textit{sleep quality tracking monitor}"). This strategic deconstruction of a general inquiry into a useful discovery journey highlights the core value of the \texttt{LESER} framework.

\input{./tab/tab4}

\subsection{Online Evaluation}
We validated \texttt{LESER}'s real-world efficacy under production latency constraints by evaluating in a large-scale A/B test. This model serves pre-computed expansions from a cache for around 600K high-frequency queries. In the 2-week experiment, we assigned 10\% of traffic to both control and treatment groups. The treatment group demonstrated statistically significant improvements in user engagement which is reflected by a higher click-through rate and a decreased in query abandonment. Consequently, these positive user signals translated into growth in overall gross merchandise value. While specific lift percentages remain confidential, the consistent improvement across the conversion funnel validates the real-world efficacy of our approach.
% We conducted online A/B testing to evaluate the effectiveness of the LESER model. The experiment was run on 10\% of live traffic, with a 6.1\% impact entity rate evenly distributed across test and control groups. The test spanned a period of two weeks and included approximately 600,000 cached queries and their corresponding query expansions. We observed consistent improvements in both the relevance of the search results and overall user engagement. However, we are unable to share detailed business metrics due to confidentiality considerations.

%% file: tab/tab2.tex
\begin{table}[t]
\centering
\small
\begin{tabular}{l|c|c|cc}
\toprule
\textbf{Model} & \textbf{Backbone} & 
$\Delta\mathsf{Ret}$ (\%) &
$\Delta\mathsf{Rel}$ (\%)\\
% \textbf{Retriv\textsubscript{Gain}} & 
% \textbf{Rel\textsubscript{Gain}} \\
\midrule
EBQE    & Sentence-BERT  & 8.99$\pm$0.00 & 26.01$\pm$0.00   \\
GPT4.1     & OpenAI Proprietary
 &  16.51$\pm$2.03   & 27.82$\pm$2.01  \\
SFT-LLM     & LLaMA 3.2-3B & 19.22$\pm$0.09  & 15.10 $\pm$1.40  \\
\midrule
 \texttt{LESER} & LLaMA 3.2-3B  & \textbf{72.84}$\pm$3.91    & \textbf{32.40}$\pm$4.37   \\
\bottomrule
\end{tabular}
\caption{Evaluation of query expansion models using retrieval gain, and relevance gain.}
\label{tab:exp1}
\end{table}

%% file: tab/tab3.tex
\begin{table}[t]
\centering
\small
\begin{tabular}{cc|c|c|c}
\toprule
\textbf{WarmUp} & \textbf{GRPO}  & \textbf{Backbone} & $\Delta\mathsf{Ret}$  (\%) & $\Delta\mathsf{Rel}$ (\%) \\
\midrule
&  &    LLaMA 3.2-3B  & 12.80$\pm$0.50 & 8.93$\pm$0.40  \\
 &  \checkmark &    LLaMA 3.2-3B  & 57.45$\pm$4.00  & 25.20$\pm$5.00  \\
\checkmark  & \checkmark &  LLaMA 3.2-1B   &  29.41$\pm$2.28   & 14.50$\pm$1.13 \\

\checkmark & \checkmark & LLaMA 3.2-3B  & 72.84$\pm$3.91    & 32.40$\pm$4.37   \\
\checkmark & \checkmark &   LLaMA 3.1-8B & 76.00$\pm$3.80   & 34.00$\pm$3.90    \\
\bottomrule
\end{tabular}
\caption{Ablation study of LESER across backbone sizes, with and without WarmUp and GRPO.}
\label{tab:exp2}
\end{table}

%% file: tab/tab4.tex
\begin{table}[t]
\centering
\small
\begin{tabular}{@{}p{\columnwidth}@{}}
\toprule
\addlinespace[0.5em]
\textbf{Query:} \textit{black bose quietcomfort ultra headphones} \\
\textbf{Generated Expansions:} None \\
\midrule

\textbf{Query:} \textit{ergonomic mouse} \\
\textbf{Generated Expansions:} \\
\quad \textbullet{} \textit{wireless vertical mouse for wrist support} \\
\quad \textbullet{} \textit{logitech mx wireless trackball} \\
\midrule
\textbf{Query:} \textit{sleep improvement} \\
\textbf{Generated Expansions:} \\
\quad \textbullet{} \textit{natural sleep aid supplement} \\
\quad \textbullet{} \textit{anti-snoring device for better breathing} \\
\quad \textbullet{} \textit{sleep quality tracking monitor} \\
\quad \textbullet{} \textit{herbal sleep relief patch} \\
\bottomrule
\end{tabular}
\caption{Qualitative examples of the LESER framework, illustrating how query ambiguity dictates the generation of expansions.}
\label{tab:qualitative_examples}
\end{table}

%% file: section/5_conclusion.tex
% \section{Conclusion}
% LESER demonstrates the value of a closed-loop system that fine-tunes language models using direct feedback from a live search engine, successfully turning ambiguous user needs into expanded and specific queries. Our use of Group Relative Policy Optimization proved highly effective, allowing the model to handle cases with multiple valid interpretations by generating a diverse set of high-quality refinements. The significant improvements over strong baselines in offline experiments, which were confirmed by a large-scale online A/B test, validate LESER's real-world efficacy and deployability in a production environment. This work paves the way for more dynamic and responsive search systems. Future work will build on this success by extending the framework to provide real-time expansions for long-tail queries.

\section{Conclusion}

In this work, we introduced \texttt{LESER}, a novel framework for context-aware query expansion in e-commerce search that learns from real-time search engine feedback via reinforcement learning (search-engine-in-the-loop). By leveraging a retrieval-informed LLM trained through GRPO, \texttt{LESER} generates expansions that are both diverse and relevant, addressing the inherent one-to-many mapping challenge posed by vague or underspecified user queries. \texttt{LESER} directly optimizes the generation process using live retrieval signals, eliminating the need for manual annotations and ensuring compatibility with platform-specific constraints. Our offline evaluations demonstrates \texttt{LESER}'s superior performance over strong baselines and online experiment showcases it significantly enhances coverage, retrieval performance, and user engagement. Looking forward, we see opportunities to extend \texttt{LESER} to more search related tasks to further boost robustness and interpretability in deployments.

%% file: section/6_appendix.tex
\clearpage
\setcounter{page}{1}
\setcounter{section}{0} % Reset the section counter
\renewcommand\thesection{\Alph{section}} % 

\section{Implementation Details}
\paragraph{Hyperparameter.} We adopt a two-stage training pipeline to align large language models with structured query generation tasks in e-commerce. All experiments are conducted using the Swift framework~\cite{zhao2024swiftascalablelightweightinfrastructure}. In the supervised fine-tuning (SFT) warm up stage, we perform full-parameter updates using \texttt{bfloat16} precision, a global batch size of 64 (with per-device batch size of 16 and gradient accumulation), a learning rate of $1 \times 10^{-7}$, and a warmup ratio of 5\%. DeepSpeed ZeRO-3 \cite{rajbhandari2020zero} is employed for memory-efficient optimization. In the GRPO training, the batch size is set to $8$ with no gradient accumulation. During the training, the model generates $N = 4$ candidates per input using temperature sampling ($\tau = 0.9$), and is optimized with a learning rate of $1 \times 10^{-6}$ with reward weight $\lambda = 0.1$ and $\beta = 0.04$. Both stages are trained on 8 NVIDIA H200 GPUs using full-parameter updates, with a maximum input length of $2048$. Furthermore, in constructing the instruction, the context set is chosen to be no more than 10 items. In the reward calculation, the relevance score is generated from internal relevance model from retrieved top 10 items, and the 

\paragraph{Relevance Model.} Following the approach in~\cite{saha2025improving}, the relevance model is based on a Gradient Boosted Decision Tree (GBDT) trained in a pointwise manner using Gaussian regression. It uses human-judged relevance labels on a five-point scale (Perfect, Excellent, Good, Fair, Bad) and incorporates incremental updates to handle new sponsored listings. The model outputs a continuous relevance score for each item comparing to the query.

\paragraph{Datasets.}
We begin by training our SFT model on 10,000 examples distilled from a general in-house dataset. For each query, we use GPT-4.1 to produce both a reasoning and a final answer using the format: \texttt{<think>Reasoning</think>} and \texttt{<answer>Answer</answer>}. The reasoning section explicitly explains why the answer is appropriate, helping inject domain-specific business logic into the model and encouraging adherence to the structured response format. The SFT model is trained for one epoch. We then continue with GRPO training. Unlike SFT, GRPO requires interaction with a search engine to compute rewards. We use a dataset of 100,000 real-world search queries, where each generated structured intent is sent to the search engine, and the resulting retrieval set is used to calculate the reward.